\documentstyle[preprint, tighten, aps, axodraw]{revtex}
\newcommand{\D}{/\! \! \! \! D}
\begin{document}
\draft                                         % REVTEX specific
\preprint{\parbox{1.5in}{\noindent UM-TH-98-11 \\ ITP-SB-98-48}} 
                                               % REVTEX specific
\title{ An operator expansion for the elastic limit }
\author{ Ratindranath Akhoury, \  Michael G. Sotiropoulos}
\address{Randall Laboratory of Physics, University of Michigan,
         Ann Arbor, MI 48109}                  % REVTEX specific
\author{George Sterman}
\address{Institute for Theoretical Physics,
         State University of New York at Stony Brook, \\  
         Stony Brook,  NY 11794}               % REVTEX specific
\date{July 10, 1998}
\maketitle
\begin{abstract}
A leading twist expansion in terms of bi-local operators is proposed for the   
structure functions of deeply inelastic scattering near the elastic limit 
$x \rightarrow 1$, which is also applicable to a range of other processes.
Operators of increasing dimensions contribute to logarithmically enhanced 
terms which are supressed by corresponding powers of $1-x$. 
For the longitudinal  structure function, in moment ($N$) space, 
all the logarithmic contributions  of order $\ln^k N/N$ are shown to be 
resummable in terms of the anomalous dimension of the leading operator in the 
expansion.
\end{abstract}
     % REVTEX specific

\narrowtext                       % REVTEX specific
\paragraph{Introduction:} 
Of special interest in QCD hard processes are ``elastic" limits,
in which the mass of the final state is small compared to
the momentum transfer.  In such limits we can study, in principle,
the transition from short-distance to long-distance dynamics.
Typically, perturbative calculations receive logarithmic enhancements
near the elastic limit.
Interesting and phenomenologically relevant examples include 
(but are not limited to):  the structure functions in deeply inelastic 
scattering (DIS) in the limit $x \rightarrow 1$ where $x=Q^2/2p\cdot q$,  
with $q^2=-Q^2$ the momentum transfer and $p$ the hadron's momentum,
the Drell-Yan cross section at partonic threshold, 
$z\equiv Q^2/\hat{s}\rightarrow 1$, with $Q$ the pair mass and 
$\hat s$ the partonic center-of-mass energy squared, and
the thrust, $T$ in ${\rm e}^+{\rm e}^-$ annihilation for $T\rightarrow 1$,
where $T\sim (m_1^2+m_2^2)/s$, with $m_i^2$ the masses
of the two final-state jets that characterize the elastic limit.

In each of these cases, logarithmic corrections at
leading power have been resummed to all orders in perturbation
theory  at leading and next-to-leading logarithmic accuracy
\cite{Sterman,CatTre,KorMar}; for
example, the terms like $\alpha_s^n [\ln^m(1-x)/(1-x)]_+$, with 
$m\le 2n-1$, in the structure function $F_2$ in DIS.
Until now, however, there has been no technique to resum
higher-order logarithmic corrections beyond the leading power of $1-T$,
$1-z$ or $1-x$.
The phenomenological importance of these corrections has been highlighted
recently  in the context of Higgs production \cite{KLS}.  They may also
play an important role
in renormalon phenomenology \cite{renormalon}.

In this paper, we take a step toward the resummation of
logarithmic corrections at nonleading powers
in the elastic limit. We propose that logarithmic
enhancements near the boundary of phase space can be analyzed systematically
through a bi-local operator product expansion. Our procedure supplements
the light-cone expansion in DIS at leading twist, by identifying new operators
associated with leading and nonleading behavior
of the coefficient functions as $x\rightarrow 1$.  It is also
applicable to processes, such as thrust, for which the light-cone
expansion is not available.
To be specific, we shall consider the example of
logarithmic enhancements to the longitudinal
structure function $F_L$, but we
expect, following our comments above, numerous other applications.

For $F_L$ power singularities at $x = 1$ are absent. 
In moment space, the leading  power terms (in $N$) are  of order 
$\alpha_s^n \ln^k N/N$. 
We shall argue that all such logarithmic contributions  can be treated 
systematically within the formalism of large-$x$ factorization  and 
the bi-local operator product expansion introduced below. 
This expansion is distinct from, although consistent with,  the light-cone 
expansion, because operators of different dimensionality all contribute to 
the leading power in $Q$, the difference being only that higher dimensional 
operators produce terms suppressed by higher powers of $1-x$.

Every inclusive DIS observable is derived as a particular projection of 
the hadronic tensor 
\begin{equation}
W_{\mu \nu}(p, q) =  \frac{1}{4 \pi}
\int d^4y \; {\rm e}^{-iq\cdot y} \; 
\langle p | j_\mu^\dagger(0) \, j_\nu(y) |p \rangle\, . 
\label{wtens}
\end{equation}
We shall denote such a four-dimensional Fourier transform, with 
respect to momentum $k$, as ${\rm FT}^{(4)}_k$ below.
For our discussion, it is enough to consider only electromagnetic 
interactions. Generalization to the electroweak case is straightforward. 
Structure functions are obtained via the projections 
$F_{\rm r} = P^{\mu \nu}_{\rm r} \, W_{\mu \nu}$, for ${\rm r} = 2, L$, with
\begin{equation}
P_L^{\mu \nu} = \frac{8 x^2}{Q^2} p^\mu p^\nu , 
\ \ \ \ \
P_2^{\mu \nu} = -\eta^{\mu\nu} +\frac{3}{2} P^{\mu\nu}_L. 
\label{projectors}
\end{equation} 

The resummation of logarithmic corrections at leading power in $F_2$ 
as $x \rightarrow 1$ may be regarded
as a consequence  of the following factorization theorem \cite{CoLaSte}:
\widetext
\begin{eqnarray}
F_2(x, Q^2) &=&  |H_2(Q^2)|^2 \, 
\int_x^1 dx' \int_0^{x'-x} dw\; J\left((x'-x-w)Q^2\right)\; V(w) \; \phi(x')
\nonumber\\
&\equiv& 
|H_2(Q^2)|^2 J \otimes V \otimes \phi\, , 
\label{f2fac}
\end{eqnarray}
\narrowtext
where $\otimes$ denotes the convolution in the longitudinal momentum 
fraction. This theorem has been proven in the axial gauge 
in Ref.~\cite{Sterman}. Below we shall define each of the factors that enter 
Eq.~(\ref{f2fac}) in a manifestly gauge independent way. 
$H_2(Q^2)$ is the short distance hard scattering function
on either side of the final state cut. 
$V$ is the soft radiation function that contains all enhancements coming 
from on-shell propagation of low frequency partons. It is defined as 
\widetext
\begin{eqnarray}
V(w) &=& \int dy^- {\rm e}^{-iw y^- p \cdot \bar v} \;
\langle 0| \Phi^\dagger_v(0, -\infty) \; 
\Phi_{\bar{v}}(0, y^- \bar v) \;  
\Phi_{v}(y^- \bar v, -\infty) |0 \rangle
\nonumber \\ 
&\equiv & {\rm FT}^{(1)}_{-wp} \; 
\langle 0| \Phi^\dagger_v(0, -\infty) \; 
\Phi_{\bar{v}}(0, y^- \bar v) \;  
\Phi_{v}(y^- \bar v, -\infty) |0 \rangle \, .
\label{Vdef}
\end{eqnarray}
\narrowtext
Here the light-like incoming direction is denoted by $v^\mu$ (+ direction) 
and the parity reflected by $\bar v^\mu$ ($-$ direction), with 
$v \cdot \bar v = 1$.
In the second line we have introduced a notation
for Fourier transforms along the light cone. 
The Wilson line operator along the $v$ light-cone direction is  
\begin{equation}
\Phi_v(x+tv, -\infty) = 
P \exp\left( -i g_s \int_{-\infty}^t d s \, 
v^\mu \, A_\mu( x+s v ) \right) \, .
\label{Phidef}
\end{equation} 
$\phi$ is the parton distribution function that contains the collinear 
enhancements from the initial state.
Its operator definition, for the scattering of quarks, is 
\begin{equation}
\phi(x') =
{\rm FT}^{(1)}_{x'p}\, \langle p| \psi(0) \not{\bar{v}} \; 
\Phi_{\bar{v}}(0,y^-) \, \bar{\psi}(y)|p\rangle \otimes V^{-1}\, , 
\label{fdef}
\end{equation} 
where $V^{-1}$ is defined by $V^{-1}\otimes V=1$.
The role of $V^{-1}$ is to remove soft contributions,
which would otherwise be double-counted in both $\phi$ and $V$. 
Finally, $J$ is a function that describes the outgoing
``current jet", of invariant mass  $(x'-x-w)Q^2$. 
The partons in $J$ are almost collinear and moving in the $\bar v^\mu$ 
light-cone direction. The operator definition of $J$ is 
\widetext
\begin{equation}
J\left((1-z)Q^2\right)= {\rm FT}^{(4)}_{q+zp}\;  \langle 0|  
\Phi^\dagger_v(0, -\infty) \; \psi(0) \, 
\bar{\psi}(y) \;  \Phi_{v}(y, -\infty) |0 \rangle \otimes V^{-1}. 
\label{jdef}
\end{equation}
\narrowtext
The Wilson lines make the functions manifestly gauge invariant. 
Whenever they do not explicitly appear in our expressions henceforth it is 
to be assumed  that they are always there.   

\paragraph{Operator expansion:}
Let us  now generalize the analysis of the leading contributions to $F_2$. 
We shall see that the method not only applies directly to $F_L$,
but also opens the way to the analysis of nonleading singular corrections in
$F_2$, and elsewhere.

As $x\rightarrow 1$, $F_L(x,Q^2)$ depends on three scales: the momentum 
transfer $Q$, the hadronic mass scale of order $\Lambda_{\rm QCD}$,  
and an intermediate scale $(1-x)^{1/2}Q$, which is essentially the mass 
of the final state.  
We assume that, although $x$ is close to unity, nevertheless 
\mbox{$(1-x)^{1/2}$} $\gg\Lambda_{\rm QCD}/Q$, so that the intermediate scale remains
perturbative.
At the same time, logarithmic behavior in $1-x$ arises from the limit in
which the final state becomes massless relative to $Q$.  
Logarithmic contributions to this limit are associated in perturbation theory
with the momentum configurations  illustrated by Fig.\ 1.  
The figure shows a cut diagram representation of the scattering process, 
in which the final state consists of a single ``jet" of particles, moving in 
the $\bar v^\mu$ direction, in the notation of the previous section.  
Following the general analysis of Ref.~\cite{CoSoSte}, in these 
regions of momentum space the lines of any diagram fall into one of four 
categories.  They are either ``soft" lines (subdiagram $S$ in the figure), 
with momenta that vanish relative to $Q$ in all four components, 
``hard" lines, which are off-shell by order $Q^2$, or ``jet-like" lines, 
with one momentum component of order $Q$, 
either in the $v^\mu$ (parallel to the incoming hadron) or $\bar v^\mu$ 
(part of the outgoing jet).  Hard lines are all contained in the subdiagrams 
labelled $H$ (in the amplitude) and $H^{\dagger}$ 
(in the complex conjugate) in Fig.\ 1. 
Note that the hard amplitudes are not specific to any structure function. 
The discussion so far is at the level of the tensor $W^{\mu\nu}$. 

Because lines in $H$ and $H^\dagger$ are off-shell by the largest momentum 
scale, $Q$, we may extend the factorization program somewhat, and treat them 
in terms of an ``effective Hamiltonian", as an expansion in products of
local fields,
\begin{equation}
H^\nu_{\rm eff}=\sum_i C_i(Q,\mu)\; O_i^\nu(0)\, ,
\label{Heff}
\end{equation}
where $\mu$ is the factorization scale and the operators $O_i^\nu$ carry the 
quantum numbers of the electromagnetic current.    
The functions $C_i$ are the usual coefficient functions that accompany
the hard-scattering function.  For the leading-power analysis of
$F_2$ \cite{Sterman}, we keep only the lowest-dimension operator, 
the original electromagnetic current $j^\nu(0)=\bar\psi(0)\gamma^\nu\psi(0)$,
the matrix elements of whose product is then factorized into the parton 
distribution $\phi$, Eq.\ (\ref{fdef}), which uses up two of the quark fields
in $j^\dagger\; j$, and the jet function $J$, Eq.\ (\ref{jdef}), which uses  
up the other two.  

Suppose that $P^{\mu\nu}_{\rm r}$ projects out one of the structure functions.
Then we can write
\widetext
\begin{eqnarray}
P^{\mu\nu}_{\rm r} W_{\mu\nu}
&=&
P^{\mu\nu}_{\rm r} \; {\rm FT}^{(4)}_{q} \; 
\langle p|\; \bar\psi(0)\gamma_\mu\psi(0)\; 
\bar\psi(y)\gamma_\nu\psi(y)\; |p\rangle
\nonumber\\
&\approx& {\rm FT}^{(1)}_{xp}
\langle p|\; \bar\psi(0)\; \psi(y^-\bar v)\; |p\rangle\, 
\otimes  \; {\rm FT}^{(4)}_{q+xp} \;
P^{\mu\nu}_{\rm r} \gamma_\mu\; \langle 0| \psi(0)\; \bar\psi(y)\; |0\rangle\;
\gamma_\nu \, .
\label{firstfact}
\end{eqnarray}
\narrowtext
Noting that the Wilson lines, Eq.\ (\ref{Phidef}), are needed to formulate 
the factorization in a gauge invariant fashion, we see that the leading order
factorization of $F_2$ is thus reproduced. It is obvious that this procedure
gives a vanishing result when the $j$'s are projected with $P^{\mu\nu}_L$ 
in Eq.\ (\ref{projectors}) to give $F_L$. 
All we have to do in this case, however, is to consider the operators of the 
next higher dimension in $H_{\rm eff}$, Eq.\ (\ref{Heff}). Upon using the
equations of motion for both $\psi$ and $\bar \psi$, the following  candidates,
all of dimension four, can be identified: 
\begin{eqnarray}
&\ &
O^\mu_{2a} = \bar v^\mu \; \bar \psi \stackrel{\leftarrow}{\D_{\perp}} \psi \, ,
 \ \ \  
O^\mu_{2b} =  v^\mu \; \bar \psi \stackrel{\leftarrow}{\D_{\perp}} \psi \, ,  
\nonumber \\ 
&\ & 
O^\mu_{2c} = \bar \psi \stackrel{\leftarrow}{D_\nu} 
\sigma^{\mu \nu}_{\perp} \psi \, ,  
\ \ \ 
O^\mu_{2d} = \bar \psi \stackrel{\leftarrow}{D^\mu} \psi \, .  
\label{candidates}
\end{eqnarray}
It is easily seen that $O_{2b}$ and $O_{2c}$ have vanishing projection when
contracted with  $P^{\mu \nu}_L$. 
They do not contribute to $F_L$ but can contribute to $F_2$ at the subleading 
level. 
$O_{2d}$ has a longitudinal projection that is
proportional to $\bar \psi \; (v \cdot \stackrel{\leftarrow}{D}) \psi$.
After integration by parts this term vanishes by the equation of motion of 
$\psi$ at $x=1$.  
We therefore conclude that the only operator of dimension four that will 
contribute to $F_L$ at leading level is $O_{2a}$ with longitudinal projection 
\begin{equation}
v_\mu O_{2a}^\mu = \bar{\psi} \;   \stackrel{\leftarrow}{\D_\perp} \psi \, . 
\label{theonlyone}
\end{equation}
Recall that the field $\psi$ will be contracted with partons from the 
incoming state and that it is 
$ \bar{\psi} \;\stackrel{\leftarrow}{\D_\perp}$ that enters the jet 
function. 
The reason that the covariant derivative is associated with
$\bar \psi$ and not on $\psi$ comes from dimensional considerations,
as explained below. 

To match the above bi-local operators to the general cut diagram in 
Fig.\ 1, we need to decide which fields inside each $O_2$ 
will be contracted with partons from the incoming state, and which with the
outgoing jet.  
A moment's thought shows that, for an incoming quark state, only the
quark field $\psi$ can be associated with the parton distribution.
This is because dimensional considerations require $C_2(Q)$ to behave as 
$Q^{-1}$, up to logarithmic corrections.  
The only way to make up this suppression, and derive a  result that is not 
power-suppressed in $Q$, is to integrate over final-state phase space.
If fields of dimension $3/2+d$ are
emitted into the final state, the resulting phase space integral behaves as
$[(1-x)Q^2]^d$, which precisely makes up for the dimensional suppression of 
$C_2$, for $d=1$, and indeed for any of the higher-dimension coefficient 
functions in the effective Hamiltonian Eq.\ (\ref{Heff}).  
No such compensation can arise when the extra fields are contracted with 
lines parallel to the initial state, because the initial state is not summed. 
Consider, then, a general operator $O_i$ in the expansion of $H_{\rm eff}$,
\begin{equation}
O_i^\mu = \Xi_i^\mu{}^\dagger\; \psi\, ,
\label{oheye}
\end{equation}
where $\Xi_i^\mu$ is some operator of dimension greater than 3/2.  
The leading power contribution of $O_i$ to $P^{\mu\nu}_{\rm r}W_{\mu\nu}$ 
as $x\rightarrow 1$ will be associated with the following factorization, 
which generalizes Eq.\ (\ref{firstfact})  directly:
\widetext
\begin{eqnarray}
P^{\mu\nu}_{\rm r}W_{\mu\nu}
&=&
P^{\mu\nu}_{\rm r} \; {\rm FT}^{(4)}_{q} \; 
\langle p|\;\bar\psi(0) \;  
\Xi_{\mu, i}(0)\; \Xi_{\nu, i}^\dagger(y) \; \psi(y)\; |p\rangle
\nonumber\\
&\approx& {\rm FT}^{(1)}_{xp} \; 
\langle p|\; \bar\psi(0)\; \psi(y^-\bar v)\; |p\rangle\, 
\otimes {\rm FT}^{(4)}_{q+xp} \; 
P^{\mu \nu}_{\rm r} \langle 0|  \; \Xi_{\mu,i}(0)\; \Xi_{\nu,i}^\dagger(y) \; 
|0\rangle \, .
\label{generalfact}
\end{eqnarray}   
\narrowtext
The second correlation function is a jet function and we see that at each 
level in the expansion of $H_{\rm eff}$ a new set of jet functions is 
generated that differ in their contributions by powers of $(1-x)$ according 
to the dimensional counting rule given above. 

\paragraph{The $F_L$ factorization:} 
Next we apply the above analysis to the $x\rightarrow 1$ 
behavior of $F_L$. Its leading behavior, which comes from  $O^\mu_{2a}$, is 
of course leading power in $Q$, but suppressed by a factor of $1-x$ compared 
to  $F_2$, which  is just what we expect. In addition, it is also 
easy to see that for a gluonic state, the lowest-order operator, is even 
higher order, so that the contribution to $F_L$ from gluons is actually 
finite in the limit $x\rightarrow 1$.
The factorization formula for the leading corrections of $F_L$ is 
\begin{equation}
F_L(x, Q^2) = |H_L(Q^2)|^2 J^\prime \otimes V \otimes \phi \, .
\label{flfac}
\end{equation} 
The factors $V$ and $\phi$ are the same as in $F_2$ case, 
see Eqs.\ (\ref{Vdef}), (\ref{fdef}). 
The new jet function $J^\prime$, generated by the 
$O_{2a}$ operator is defined by 
\widetext
\begin{equation}
J^\prime((1-z) Q^2) = 
 \left(\frac{1}{4\pi} \; \frac{8 x^2}{Q^2} \right) \; 
 {\rm FT}^{(4)}_{q+zp} \;  
\langle 0| \Phi_{v}^\dagger (0, -\infty) \; \D_\perp \psi(0) \; 
\D_\perp\bar{\psi}(y) \; \Phi_v (y, -\infty) |0 \rangle \otimes V^{-1}.
\label{Jprimedef}
\end{equation}
\narrowtext
The conventional normalization factors of 
Eqs.~(\ref{wtens}), (\ref{projectors}) have been included. 
Once the definition is given in terms of an effective operator, 
the anomalous dimension of $J^\prime$ can be computed in dimensionally 
regularized perturbation theory with massless partons.
Details of this calculation will be published in a forthcoming paper. 
The following points, however, are worth noting here: First, one has to be careful to
distinguish between  the UV renormalization  of the operator in $J^\prime$ 
that corresponds to  genuine collinear enhancements, from the renormalization 
of the coupling constant. This is as easy task since the renormalization 
of the coupling is well known. Second, since  $J^\prime$ starts at
${\cal O}(\alpha_s)$ whereas $J$,  Eq.\ (\ref{jdef}), starts at
${\cal O}(\alpha_s^0)$,  the running coupling will mix 
with the anomalous dimension of $J^\prime$ already at the first non-trivial 
order ${\cal O}(\alpha_s^2)$. 

The resummation of the leading corrections to $F_L$ is a consequence of its  
factorization. First, the factorization formula is written in moment space 
in terms of the Mellin transformed factors:
\widetext
\begin{eqnarray}
\tilde{F}_L(N, Q^2, \epsilon) &=& 
\left|H_L\left(\frac{(p\cdot \bar{v})^2}{\mu^2},
\frac{(\bar{p}\cdot v)^2}{\mu^2},\alpha_s(\mu^2) \right)\right|^2 
\nonumber \\ 
&\ & \hspace{-2.5cm} \times \frac{1}{N} 
\tilde{J}^\prime \left( \frac{Q^2}{N \mu^2}, 
\frac{(\bar{p} \cdot v)^2}{\mu^2}, \alpha_s(\mu^2) \right) \, 
\tilde{V}\left( \frac{Q^2}{N^2 \mu^2},  \alpha_s(\mu^2)\right) \, 
\tilde{\phi}\left( \frac{Q^2}{N^2 \mu^2}, \frac{(p \cdot \bar{v})^2}{\mu^2}, 
\alpha_s(\mu^2), \epsilon \right) \, ,   
\label{flfacconv}
\end{eqnarray}  
with $\bar{p}^\mu$ the jet momentum at the elastic limit.
The Mellin transformation of the jet function $J^\prime$ 
is defined without including the overall $1/N$, 
\begin{equation}
\frac{1}{N} 
\tilde{J}^\prime\left( \frac{Q^2}{N \mu^2}, \frac{(\bar{p} \cdot v)^2}{\mu^2},
\alpha_s(\mu^2)\right) 
= \int_0^1 dz \, z^{N-1} \, 
J^\prime \left(\frac{(1-z) Q^2}{x \mu^2}, \frac{(\bar{p} \cdot v)^2}{\mu^2}, 
 \alpha_s(\mu^2) \right) \, . 
\label{Mellinj}
\end{equation}
\narrowtext
The factor of $1/N$ arises because, as seen from the dimensional counting 
of the previous section, the jet $J^\prime$ is of order $1-x$ relative to 
the $F_2$ jet $J$. It is this additional factor of $1-x$ that shows up as 
a power suppresion $1/N$ in moment space. Once this is isolated,  
the renormalization group can be used to resum the logarithmic in $N$ terms. 

The resummation formula of the logarithmic corrections is derived using 
the, by now standard, methods that are applicable to a large range of 
processes near their elastic limits. 
Enhancements from soft gluon radiation exponentiate. 
The evolution of the incoming jet $\phi$ can also be written as an exponential.
These two factors are common to both $F_2$ and $F_L$ and we shall not 
reproduce them here (for review and details see Ref.~\cite{CoLaSte}).
We concentrate on the differences between the two strucutre functions at 
leading power in $N$.
The renormalization group equation satisfied by the jet function $J^\prime$ is 
\begin{equation} 
\frac{d}{d \ln \mu^2} 
\ln \tilde{J}^\prime \left(\frac{Q^2}{N \mu^2}, 
\frac{(\bar{p}\cdot v)^2}{\mu^2}, \alpha_s(\mu^2) \right) 
= -\frac{1}{2}\gamma_{J^\prime}(\alpha_s(\mu^2)) \, . 
\label{rg}
\end{equation}
The anomalous dimension $\gamma_{J^\prime}$ is novel and characteristic 
of $F_L$ at the leading power level. 
Up to corrections of ${\cal O}(\alpha_s^2)$ it turns out to be \cite{ASS} 
\begin{equation}
\gamma_{J^\prime}(\alpha_s) = \frac{\alpha_s}{\pi} 
\left[ \frac{9}{2}C_F -2 C_A - 4 \zeta(2) \left(C_F-\frac{C_A}{2}\right)
\right] . 
\label{gammaJtotno}
\end{equation}
The initial condition for Eq.\ (\ref{rg}) follows from the lowest order 
contribution to $F_L$, which comes entirely from $J^{\prime}$. It is 
\begin{equation}
\tilde{J}^\prime(\alpha_s(\mu^2)) = C_F \frac{\alpha_s(\mu^2)}{\pi}
+ {\cal O}(\alpha_s^2(\mu^2)) \, .
\label{initJprime}
\end{equation}

The final result for $\tilde{F}_L$ may be written 
in a variety of exponentiated forms depending on which anomalous dimensions 
we wish to include in the exponent  and what initial conditions we choose 
for the factors $H_L$, $\tilde{J}^\prime$, $\tilde{V}$ and $\tilde{\phi}$. 
All these equivalent forms resum the logarithmic corrections at leading 
power, i.e. all terms of order $\ln^kN/N$.
Furthermore, we can exploit the fact that apart from their respective 
final state jets, $F_2$ and $F_L$ are similar. This means that the moments 
of the coefficient functions in the light-cone expansion  $\tilde C_2$ and 
$\tilde C_L$ are connected in a precise way, i.e. they differ only in terms 
that depend on $\gamma_J$ and $\gamma_{J^\prime}$ respectively. 
In dimensionally regularized perturbation theory this relation is
\begin{equation}
\tilde C_L = \frac{\tilde F_L(N, Q^2, \epsilon)}{\tilde F_2(N, Q^2, \epsilon)}
\tilde C_2 \, .
\label{link}
\end{equation}
The ratio of the two structure functions is infrared finite and can be 
obtained from their exponentiated forms. 
A consequence of Eqs.~(\ref{initJprime}) and (\ref{link}) is that, given 
$\tilde C_2$ to some order in $\alpha_s$, we can readily predict $\tilde C_L$ 
to one order higher in $\alpha_s$. 
To ${\cal O}(\alpha_s^2)$, using the result in Eq.~(\ref{gammaJtotno}),
we obtain in the $\overline{\rm MS}$ scheme
\begin{eqnarray}
&\ & \tilde C_L  \; = \;  \frac{\alpha_s}{\pi} \frac{C_F}{N} 
\nonumber \\
&\ & 
+\left(\frac{\alpha_s}{\pi}\right)^2 \frac{C_F}{2 N} 
\left( \gamma_K^{(1)} \ln^2 \frac{N}{N_0} - (\gamma_{J^\prime}^{(1)}
-\frac{1}{2}\beta_1)\ln \frac{N}{N_0}
\right)  , 
\label{res}
\end{eqnarray}
with $\gamma_K$ the cusp anomalous dimension \cite{Sterman,CatTre,KorMar} 
and $N_0=e^{-\gamma_E}$.
This is in agreement with the full ${\cal O}(\alpha_s^2)$ calculation of 
$C_L$ \cite{SG,ZvN}.

In conclusion, we emphasize that the operator product expansion presented in
this letter can be used to provide a systematic treatment of the logarithmic 
enhancements at the edges of phase space at both leading and subleading 
powers in moment space for a large number of hard processes in QCD. 
Further details and applications will be discussed in future publications.

\paragraph{Acknowledgements:} We would like to thank the Aspen Center
for Physics for its hospitality during the early
stages of this project.  The work of A.R. and M.G.S. was supported in part 
by the US Department of Energy. The work of G.S. was supported in part
by the National Science Foundation, grant PHY9722101.

                     % REVTEX specific 

\begin{figure}
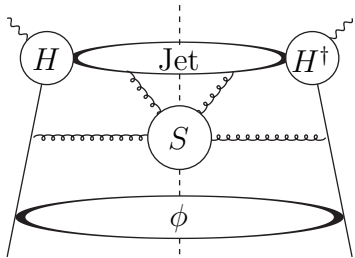

\caption{Momentum configurations that produce logarithmic enhancements 
near the elastic limit $x \rightarrow 1$ in DIS.}
\label{fig1}
\end{figure}

%--------------------- FIGURE ------------------------------

\begin{center}
\begin{picture}(200, 150)
\SetOffset(20,0)
%---------------- Cut -------------------
\DashLine(80,140)(80,125){2}
\DashLine(80,114)(80,68){2}
\DashLine(80,52)(80,46){2}
%------------ The soft subgraph ----------
\Gluon(25,90)(80,90){1}{14}
\Gluon(80,90)(135,90){-1}{14}
\Gluon(60,115)(80,90){1}{7}
\Gluon(100,115)(80,90){1}{7}
\BCirc(80,90){12}
\Text(80,90)[]{$S$}
%--------------- Jet function ---------
\Oval(80,120)(6,40)(0)
\Text(80,120)[]{Jet}
%---------------- Hard subgraphs ----------
\Photon(15,135)(30,120){1}{5}
\Photon(145,135)(130,120){1}{5}
\Line(15,45)(30,120) 
\Line(145,45)(130,120)
\BCirc(30,120){10}
\BCirc(130,120){10}
\Text(30,120)[]{$H$}
\Text(130,120)[]{$H^\dagger$}
%-------------- the PDF --------
\Oval(80, 60)(8, 60)(0) 
\Text(80, 60)[]{$\phi$}
\end{picture}
\end{center}

\begin{references}                   %  REVTEX specific
\bibitem{Sterman} G.\ Sterman, Nucl.\ Phys.\ {\bf B281}, 310 (1987).  
\bibitem{CatTre}  S.\ Catani and L. Trentadue, Nucl.\ Phys.\ {\bf B327}, 323 
  (1989); {\bf B353}, 183 (1991); L.\ Magnea,
  Nucl.\ Phys.\ {\bf B349}, 703 (1991). 
\bibitem{KorMar} G.P.\ Korchemsky and G.\ Marchesini, 
        Phys.\ Lett.\ B \ {\bf 313}, 433 (1993).
\bibitem{KLS} 
  M.\ Kr\"{a}mer,  E.\ Laenen and M.\ Spira, Nucl.\ Phys.\ {\bf B511}, 523 (1998).
\bibitem{renormalon} R.\ Akhoury and V.I.\ Zakharov, Nucl.\ Phys.\ (Proc. Suppl.)
    {\bf B54A}, 217 (1997);
    S.\ Catani, rep. no. CERN-TH/97-371, hep-ph/9712442.
\bibitem{CoLaSte}  H.\ Contopanagos, E.\ Laenen, and G.\ Sterman, 
          Nucl.\ Phys.\ {\bf B484}, 303, (1997). 
\bibitem{CoSoSte} J.C.\ Collins, D.E.\ Soper and G.\ Sterman, in 
{\it Perturbative Quantum Chromodynamics}, edited by A.H.\ Mueller,
 World Scientific, Singapore, 1989, p.\ 1. 
\bibitem{ASS} R.\ Akhoury, M.G.\ Sotiropoulos and G.\ Sterman, in preparation. 
\bibitem{SG} J.\ S\'{a}nchez Guill\'{e}n {\it et al.}, 
 Nucl.\ Phys.\ {\bf B353}, 337 (1991).
\bibitem{ZvN}  E.B.\ Zijlstra and W.L.\ van Neerven, 
              Nucl.\ Phys.\ {\bf B383}, 525 (1992). 
\end{references}
\end{document}